\documentclass[conference]{IEEEtran}
\IEEEoverridecommandlockouts
\usepackage{cite}
\usepackage{amsmath,amssymb,amsfonts}
\usepackage{algorithmic}
\usepackage{graphicx}
\usepackage{textcomp}
\usepackage{xcolor}
\usepackage{rotating}
\usepackage{multirow}
\usepackage[inline]{enumitem}
\usepackage{color,soul}
\usepackage{amssymb}
\usepackage{pifont}
\newcommand{\cmark}{\ding{51}}%
\newcommand{\xmark}{\ding{55}}%
\def\BibTeX{{\rm B\kern-.05em{\sc i\kern-.025em b}\kern-.08em
    T\kern-.1667em\lower.7ex\hbox{E}\kern-.125emX}}

\renewcommand{\theenumi}{\roman{enumi}}%
\usepackage{todonotes}

\begin{document}
\title{Towards Synthesizing Datasets for\\ IEEE 802.1 Time-sensitive Networking}


\author{
\IEEEauthorblockN{
{Do\u{g}analp Ergen\c{c}\IEEEauthorrefmark{1},
Nurefşan Sertbaş Bülbül\IEEEauthorrefmark{1},
Lisa Maile\IEEEauthorrefmark{2}, 
Anna Arestova\IEEEauthorrefmark{2},  
Mathias Fischer \IEEEauthorrefmark{1}}
}
\IEEEauthorblockA{
University of Hamburg\IEEEauthorrefmark{1}
University of Erlangen-Nürnberg \IEEEauthorrefmark{2}
\\
Email: \IEEEauthorrefmark{1}name.surname@uni-hamburg.de
\IEEEauthorrefmark{2}name.surname@fau.de}
}

\maketitle 
\begin{abstract} 
IEEE 802.1 Time-sensitive Networking~(TSN) protocols have recently been proposed to replace legacy networking technologies across different mission-critical systems~(MCSs). Design, configuration, and maintenance of TSN within MCSs require advanced methods to tackle the highly complex and interconnected nature of those systems. Accordingly, artificial intelligence~(AI) and machine learning~(ML) models are the most prominent enablers to develop such methods. However, they usually require a significant amount of data for model training, which is not easily accessible. This short paper aims to recapitulate the need for TSN datasets to flourish research on AI/ML-based techniques for TSN systems. Moreover, it analyzes the main requirements and alternative designs to build a TSN platform to synthesize realistic datasets.
\end{abstract}


\section{Introduction} 
Modern mission-critical systems~(MCSs) such as avionics and automobiles have evolved from static and close-loop networks of embedded devices to highly interconnected network of different services. IEEE 802.1 Time-sensitive Networking~(TSN) protocols have recently been proposed to satisfy the varying quality of service~(QoS) and reliability requirements of such services over standard Ethernet equipment~\cite{ieeetsn}. Replacing multiple domain-specific networking technologies in MCSs, TSN reduces their design and maintenance cost and offers an additional configuration flexibility.

This flexibility enables us (re)configuring data streams for dynamically changing data traffic requirements due to the addition or removal of new devices, mobility of existing components, or any anomalies in case of potential failures and security incidents~\cite{maile_online_admission}. It requires capturing the complex system behavior to detect such changes and develop advanced reconfiguration strategies that can adapt MCSs in real-time to ensure end-to-end deterministic communication requirements~\cite{sertbas21lcn}.

Artificial intelligence~(AI) and machine learning~(ML) have been recently employed to model the complex nature of MCSs to enhance their safety, reliability, and efficiency~\cite{Laplante2020}. 
In TSN-enabled MCSs, they can \textit{autonomously classify different types of TSN traffic} such as video streaming, event- or time-triggered traffic. This classification capability aids in network management and QoS optimization to develop effective self-configuration mechanisms~\cite{sertbas21lcn}. 
Moreover, AI/ML enables \textit{prediction of future traffic behavior}, which is crucial for capacity planning, resource allocation, network optimization, and predictive maintenance. This leads to better resource efficiency and QoS in highly dynamic environments by rearranging resources based on estimated traffic patterns~\cite{sertbas22noms}.
Besides, combined with an effective TSN monitoring tool~\cite{zeek}, AI/ML models can accurately \textit{detect anomalies in both time-sensitive data traffic}, which could indicate network intrusions, security breaches, or performance issues.

However, AI/ML models usually require training with significant amounts of data that should accurately reflect network topology and communication. 
Since IEEE 802.1 TSN protocols have yet to be broadly deployed, it is challenging to find actual data.
Besides, the lack of transparency in MCSs due to their safety and security obligations prevents even (potentially) existing data from being publicly available. 
Therefore, we need reliable sources and platforms to obtain extensive and realistic TSN datasets. 
Accordingly, in this short paper, our first goal was to recapitulate the need for TSN datasets to flourish research on AI/ML-based TSN design, configuration, and resilience methods. In the remainder, we explore the main requirements of an open and reusable platform to synthesize public TSN datasets (Section~\ref{sec:req}). Then, we shortly review alternative designs to build such a platform~(Section~\ref{sec:method}).

\section{Platform Requirements for Dataset Synthesis} \label{sec:req}

The two primary reasons for the absence of public TSN datasets are the lack of widely-deployed TSN systems and the opaque nature of MCSs. 
Designing an accessible TSN platform, e.g., a TSN-based prototype, simulator, etc., should be the first step to synthesizing the required datasets, which should represent the overall behavior of the respective system in a reliable and verifiable manner. Accordingly, we outline the requirements of such a platform as follows.

\begin{itemize}[leftmargin=*] 
\item \textbf{Representation of a realistic MCS:} A TSN-based platform (and its respective dataset) should accurately reflect the design and operational principles of actual MCSs such as aircraft, automobiles, or industrial systems. This includes modeling a realistic network topology and different service classes with distinct QoS and reliability requirements. 

\item \textbf{Support for various TSN protocols:} This platform should support a broad set of TSN protocols to generate extensive datasets reflecting various MCS scenarios. For instance, P802.1Qav CBS or P802.1Qbv TAS could be alternatively used for scheduling mixed-criticality streams. P802.1CB FRER is required for redundant communication, and P802.1Qcc SRP could also be a prerequisite for the network-wide configuration of these protocols. 

\item \textbf{Verification of system behavior:} The design and configuration of the TSN platform should be verified to ensure a reliable dataset reflecting the desired network behavior. For instance, verifying the TAS scheduler guarantees that all streams in the synthesized dataset are realistically forwarded within their latency boundaries~\cite{arestova2020design}. This requires an extensive analysis of the resulting dataset.

\item \textbf{Visibility of individual components:} The platform should allow extracting local traffic from selected components alongside an extensive dataset with system-wide network communication. This enables the analysis of the behavior in target components more isolatedly.

\item \textbf{Scalability:} A realistic topology size and connectivity of MCSs should also be considered for the platform design since they shape the interdependencies between TSN components. This mainly affects both amount and depth of a TSN dataset extracted from the respective platform.
\end{itemize}

\section{Platform Design} \label{sec:method}

We consider three potential designs for a TSN platform to synthesize datasets: a hardware-based prototype, a hybrid emulation, and a simulation platform. This section briefly analyzes them regarding the requirements mentioned earlier. Table~\ref{tab:comparison} also summarizes this discussion.

\begin{table}[h!]
\centering
 \label{tab:comparison}
 \caption{Comparison of different platform designs.}
 \begin{tabular}{c||c|c|c} 
 & \textbf{Hardware} & \textbf{Hybrid} & \textbf{Simulation}  \\ \hline\hline
 Representation & \cmark & \cmark & $\sim$   \\ 
 Protocol Support & $\sim$ & $\sim$ &  \cmark  \\
 Verification & \cmark & \cmark & \cmark   \\ 
 Visibility & $\sim$ & \cmark & \cmark   \\
 Scalability & \xmark & $\sim$ &  \cmark  \\ \hline
 \end{tabular}
\end{table}

\paragraph{Hardware-based Prototype}
A platform of actual off-the-shelf equipment, e.g., TSN bridges or more generic TSN-supporting Linux devices, represents an MCS the most realistically~\cite{Bosk2022}. While generic equipment provides extensive visibility, TSN bridges require monitoring capabilities such as mirror ports, which may only exist in some commercial TSN bridges. Proportional to their visibility, the behavior of hardware-based components can be verified by manually accessing these components or processing the dataset derived from the platform. Unfortunately, such a platform potentially has a \textit{partial} protocol support (marked as "$\sim$" in Table~\ref{tab:comparison}) since existing TSN equipment implements a limited set of protocols as it takes a significant engineering effort. Besides, it can only have a limited scalability that may not reflect the typical size and complexity of real MCSs~\cite{ergenc23demo}.

\paragraph{Hybrid Emulation}
When hardware capabilities and platform scalability are limited, it is possible to emulate certain parts of a platform by integrating software-based solutions, e.g., an emulator, into a basis hardware-based prototype. Mininet, for instance, could represent a network of generic Linux-based devices with TSN scheduling capabilities~\cite{ulbricht2022tsnflextest}. While this is partially scalable and provides more visibility and easier verification, integrating hardware- and software-based solutions is technically challenging and requires the synchronization of different environments. 

\paragraph{Simulation}
Simulation platforms have already been used in several TSN studies~\cite{sertbas21lcn, sertbas22noms}, and it is the most flexible alternative regarding configurability, scalability, and visibility. They offer a wide range of TSN protocols with a convenient level of abstraction~\cite{Steinbach2011}. Besides, integrating (even run-time) verification mechanisms into the simulations is more straightforward. However, they can only approximate the actual dynamics of MCSs and potentially deviate from real-world network behavior.

\section{Conclusion}
AI/ML models can significantly foster the development of advanced design, configuration, and maintenance techniques for emerging IEEE 802.1 TSN protocols. While these models typically require a significant amount of data for training, there does not exist any public TSN dataset due to the lack of broadly-deployed TSN systems. In this paper, we first recapitulate the need for an extensive TSN dataset to enable AI/ML research providing a standardized and reproducible basis for experimentation and validation. However, this first requires a realistic TSN platform that synthesizes and validates TSN datasets. Therefore, we next outline five major requirements and analyze alternative designs of such a platform. We aim to use this preliminary study as a road map for our ongoing effort to create extensive, realistic, and public TSN datasets.

\bibliographystyle{ieeetr}
\bibliography{IEEEabrv,references}
\end{document}